\documentstyle[multicol,pre,aps,epsf]{revtex}
\begin{document}
\draft
\title{The Link Overlap and Finite Size Effects for the 3D Ising Spin Glass}

\author{Barbara Drossel, Hemant Bokil, M.A. Moore and A.J. Bray
} 
\address{Theory Group, Department of Physics and Astronomy, University of
Manchester, Manchester M13 9PL, UK} 
\date{\today} 
\maketitle

\begin{abstract}
We study the link overlap between two replicas of an Ising spin glass
  in three dimensions using the Migdal-Kadanoff approximation and
  scaling arguments based on the droplet picture.  For moderate system
  sizes, the distribution of the link overlap shows the asymmetric
  shape and large sample-to-sample variations found in Monte Carlo
  simulations and usually attributed to replica symmetry
  breaking. However, the scaling of the width of the distribution, and
  the link overlap in the presence of a weak coupling between the two
  replicas are in agreement with the droplet picture. We also
  discuss why it is impossible to see the asymptotic droplet-like
  behaviour for moderate system sizes and temperatures not too far below the 
critical temperature.

\end{abstract}
\begin{multicols}{2}

\section{Introduction}

A series of computer simulations performed in the past few years
\cite{parisi98a,parisi98b} appear to support the claim that the
three-dimensional Edwards-Anderson spin glass shows signatures of
replica-symmetry breaking (RSB), implying the existence of infinitely
many pure states in the low-temperature phase.  In contrast, almost
rigorous arguments \cite{newman97} and recent experiments \cite{jon98}
favor the droplet picture \cite{McMillan84,fisher86,bray87} with only
one pair of pure states.

In a recent paper \cite{moo98}, we have suggested that due to the high
temperatures and small system sizes the computer simulations
\cite{parisi98a,parisi98b} are strongly affected by the critical point
and do not reflect the true low-temperature behaviour. This suggestion
was supported by a numerical calculation of the Parisi overlap
function using the Migdal-Kadanoff approximation (MKA). For the system sizes
and temperatures typically used in computer simulations we found
overlap functions similar to those in \cite{parisi98a}, however for
lower temperatures we found agreement with the predictions of the
droplet picture. That the results of computer simulations are strongly
affected by the critical point can also be concluded from
\cite{berg98}, where the Parisi overlap function shows critical
scaling (with effective exponents) down to temperatures $\simeq 0.8
T_c$. 

In recent publications \cite{mar98,mar98a,mar99}, it is claimed that
nontrivial behaviour of a quantity called the link overlap is a
reliable indicator of RSB. However, in order to place such a claim on
solid ground, one would have to show that the data cannot be
interpreted within the framework of the droplet picture.  It is the
goal of this paper to furnish this discussion which has been missing
so far. As in \cite{moo98}, we use the MKA which is known to agree
with the droplet picture. Our results show, as in the case of the
Parisi overlap function, that several nontrivial features attributed
to RSB are in fact due to finite-size effects, and that the numerical
data on the link overlap published so far are indeed in agreement with
the droplet picture. We also derive expressions for the effective 
coupling at any temperature as a function of system size and find
that one indeed needs rather large systems or low temperatures
to see droplet-like behaviour.

The outline of this paper is as follows: After introducing the model
and defining the quantities to be evaluated, we present first our
analytical and numerical results for the link overlap distribution
function. Then, we evaluate  the link overlap in the presence of a weak
coupling between the two replicas. In the following section we
explain why finite size effects are so large for the three--dimensional
Ising spin glass. Finally, we summarize and discuss
our findings.

\section{Definitions} 

The Hamiltonian $H_0$ of the Edwards-Anderson (EA) spin glass in the absence of 
an
external magnetic field is given by 
$$ \beta H_0=-\sum_{\langle i,j\rangle} J_{ij} \sigma_i\sigma_j,$$ where 
$\beta=1/k_BT$. The
Ising spins can take the values $\pm 1$, and the nearest-neighbour
couplings $J_{ij}$ are independent from each other and gaussian
distributed with a standard deviation $J$.

It has proven useful to consider two identical copies (replicas) of
the system, and to measure overlaps between them. This gives
information about the structure of the low-temperature phase, in
particular about the number of pure states. The quantity considered in this 
paper is the link overlap 
\begin{equation}
q^{(L)}(\epsilon)=(1/N_L)\sum_{\langle i,j\rangle}\langle
\sigma_i \sigma_{j} \tau_i \tau_{j}\rangle \label{ql}
\end{equation}
where the sum is over all nearest-neighbour pairs $\langle i,j\rangle$ of a lattice with $N_L$ bonds and $N$ sites, and the brackets denote the thermal and disorder average. $\sigma$ and
$\tau$ denote the spins in the two replicas. The Hamiltonian used for the
evaluation of the thermodynamic average is 
\begin{equation}
\beta H[\sigma,\tau] = \beta H_0 [\sigma]+\beta H_0 [\tau]- \epsilon\sum_{\langle i,j\rangle}
\sigma_i \sigma_{j} \tau_i \tau_{j}\, , \label{H}
\end{equation}
where $H_0$ is the ordinary spin glass Hamiltonian given above, and
the term in $\epsilon$ introduces a coupling between the two replicas.

In cases where the random couplings $J_{ij}$ are  taken to have the values $\pm 
1$, the link overlap is identical to the energy overlap. 
The main qualitative differences between the Parisi overlap 
$$
q^{(P)}=\sum_{i=1}^{N} (1/N)\langle \sigma_i \tau_i\rangle\, ,
$$
and the
link overlap 
are (i) that flipping all spins in one of the two
replicas changes the sign of $q^{(P)}$ but leaves $q^{(L)}$ invariant,
and (ii) that flipping a droplet of finite size in one of the two
replicas changes $q^{(P)}$ by an amount proportional to the volume of
the droplet, and $q^{(L)}$ by an amount proportional to the surface of
the droplet. 

Below, we will show that, just as for the Parisi overlap, the MKA can
reproduce all the essential features of the link overlap found in
Monte Carlo simulations. These results refute the claim made in
\cite{mar99} that the agreement between the MKA and  simulations
for the Parisi overlap reported in \cite{moo98} is a mere coincidence
that does not extend to the link overlap. The conclusion must be drawn
that there is no evidence for RSB in three dimensional Ising spin
glasses. 

\begin{figure}
\centerline{
\epsfysize=0.09\columnwidth{\epsfbox{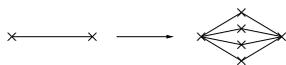}}}
\narrowtext{\caption{Construction of a hierarchical lattice.}
\label{fig0} 
}
\end{figure}

Evaluating a thermodynamic quantity in MKA in three dimensions is
equivalent to evaluating it on a hierarchical lattice that is
constructed iteratively by replacing each bond by eight bonds, as
indicated in Fig.~\ref{fig0}. The total number of bonds after $I$
iterations is $8^I$, which is identical to the number of lattice sites
of a three-dimensional lattice of size $L=2^I$.  Thermodynamic
quantities are then evaluated iteratively by tracing over the spins on
the highest level of the hierarchy, until the lowest level is reached
and the trace over the remaining two spins is calculated
\cite{southern77}. This procedure generates new effective couplings,
which have to be included in the recursion relations. In
\cite{gardner84}, it was proved that in the limit of infinitely many
dimensions (and in an expansion away from infinite dimensions) the MKA
reproduces the results of the droplet picture. We have shown in
\cite{moo98} that the MKA agrees with the
droplet picture in three dimensions as well. For this reason, 
no feature that is seen in MKA
can be attributed to RSB. 

\section{The probability distribution of the link overlap}

We first set the coupling strength $\epsilon$ in Eq.~(\ref{H}) to zero
and study the probability distribution $P(q^{(L)})$ of the link
overlap, averaged over a sufficiently large number of samples. RSB
should manifest itself in $P(q^{(L)})$ according to
\cite{mar98,mar98a} in an asymmetric (non-Gaussian) shape and a
nonzero width even at infinitely large system sizes. Furthermore, the
link overlap for single samples should show large variations between
different samples. Here, we show that the asymmetric shape and large
sample to sample variations can even be seen in MKA for moderate
system sizes and can therefore not be taken as evidence for RSB. The
only reliable indicator for RSB would be a width of $P(q^{(L)})$ that
does not shrink with increasing system size. However, the only
Monte Carlo simulation data published so far for $P(q^{(L)})$
\cite{mar98a} are taken for a four-dimensional Ising spin glass in a magnetic 
field, and they show a shrinking width, in agreement with the
expectations from the droplet picture. 

We have obtained the function $P(q^{(L)})$ by first calculating its Fourier 
transform,
$$
F(y)=\left< \exp\left(iy\sum_{\langle i j\rangle}
{\sigma_i \tau_i \sigma_j \tau_j \over {N_L}}\right) \right> .
$$
The coefficients $a_n$ in
$$
P(q^{(L)})=\sum_{n=-N_L/2}^{N_L/2} a_n \delta(q^{(L)}-2n/N_L).
$$
are then found from $F(y)$:
$$
a_n=(1/\pi N_L) \int_{-\pi N_L/2}^{\pi N_L/2} F(y) \exp(2iyn/N_L) dy.
$$

\begin{figure}
\epsfysize=0.7\columnwidth{{\epsfbox{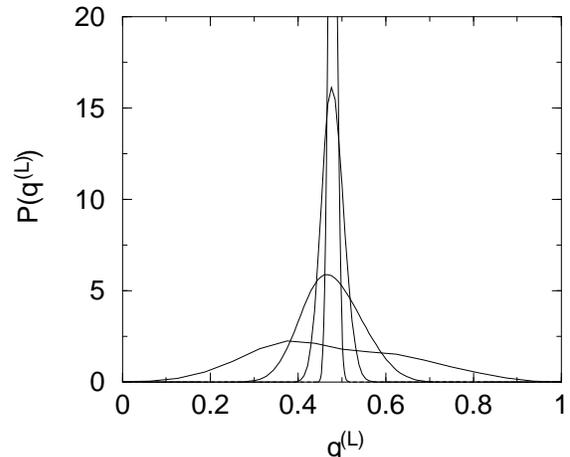}}}
\narrowtext{\caption{The link overlap distribution $P(q^{(L)})$ for $T=T_c$ and 
$L=$ 4,8,16,32 (from widest to narrowest curve). 
}
\label{fig1} }
\end{figure}

\begin{figure}
\epsfysize=0.7\columnwidth{{\epsfbox{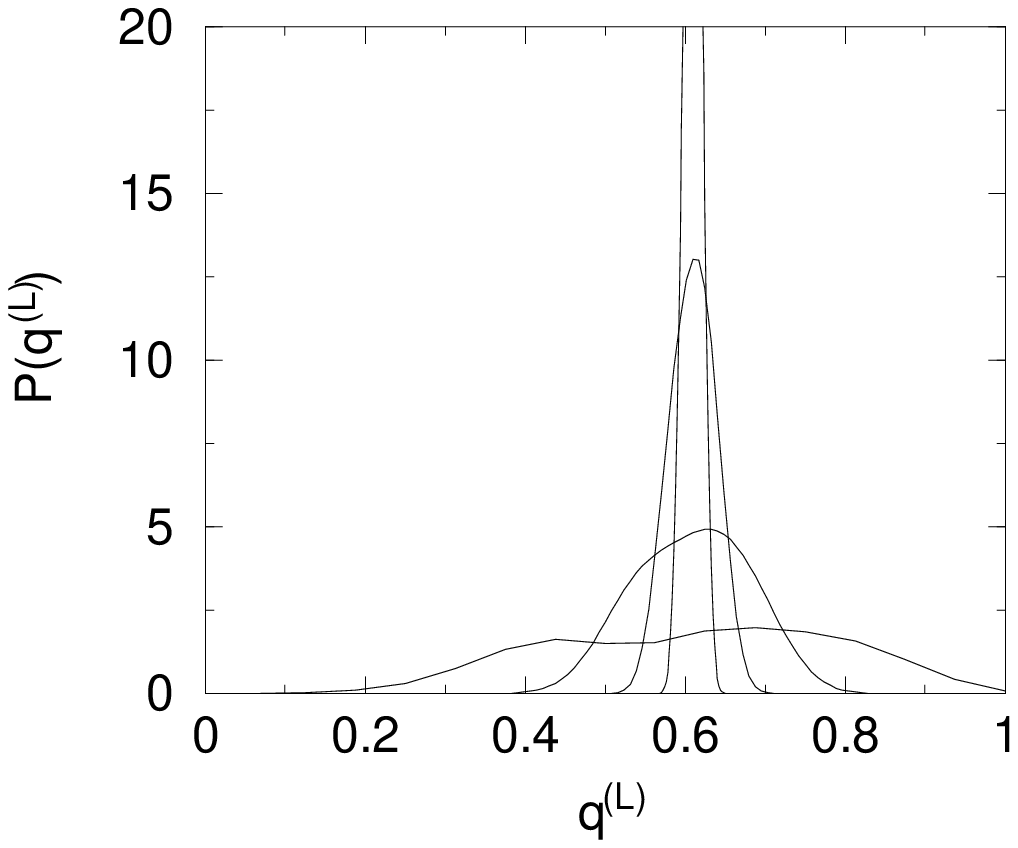}}}
\narrowtext{\caption{The link overlap distribution $P(q^{(L)})$ for $T=0.7T_c$ 
and $L=$ 4,8,16,32 (from widest to narrowest curve). 
}
\label{fig2} }
\end{figure}

\begin{figure}
\epsfysize=0.7\columnwidth{{\epsfbox{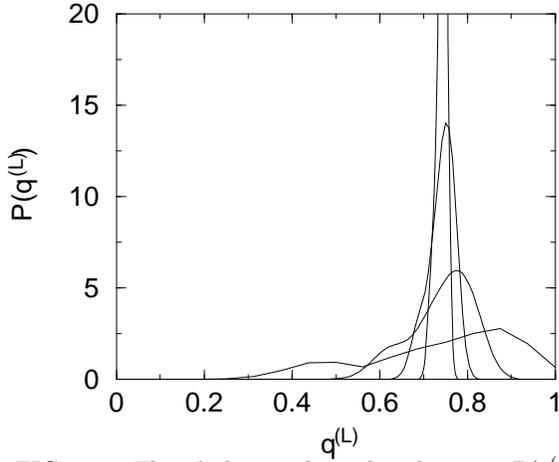}}}
\narrowtext{\caption{The link overlap distribution $P(q^{(L)})$ for
$T=0.45T_c$ and $L=$ 4,8,16,32 (from widest to narrowest curve).  }
\label{fig3} }
\end{figure}

Figures \ref{fig1}, \ref{fig2} and \ref{fig3} show our result for $P(q^{(L)})$ 
in MKA for three different temperatures. All curves have been averaged over 
several thousand samples. 

The curves for small $L$ are asymmetric at $T=T_c$ with a tail on the
right-hand side. With decreasing temperature, the asymmetry becomes
stronger, the tail moves to the left-hand side, and a shoulder is
formed. All these features seem to be finite-size effects, as they
become weaker with increasing system size.

\begin{figure}
\epsfysize=0.7\columnwidth{{\epsfbox{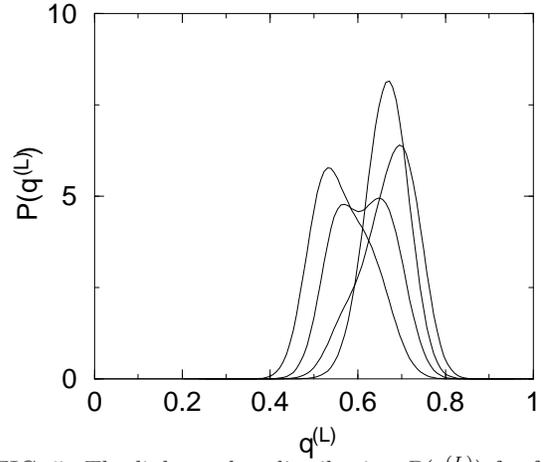}}}
\narrowtext{\caption{The link overlap distribution $P(q^{(L)})$ for
four different samples at $L=8$ and $T=0.7T_c$.  }
\label{fig4} }
\end{figure}

\begin{figure}
\epsfysize=0.7\columnwidth{{\epsfbox{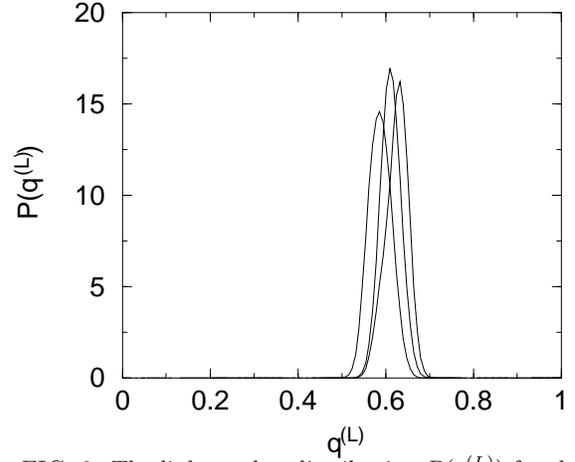}}}
\narrowtext{\caption{The link overlap distribution $P(q^{(L)})$ for
three different samples at $L=16$ and $T=0.7T_c$.  }
\label{fig5} }
\end{figure}

Figures \ref{fig4} and \ref{fig5} show $P(q^{(L)})$ for single samples
at $T=0.7T_c$ and for $L=8$ (Fig.~\ref{fig4}) and $L=16$
(Fig.~\ref{fig5}). 
In particular for $L=8$, there are large variations between different
samples, a feature that is usually assumed to be a clear indicator of
RSB. However, since the MKA does not show RSB,
we must asign this feature to
finite-size effects. 

\begin{figure}
\epsfysize=0.7\columnwidth{{\epsfbox{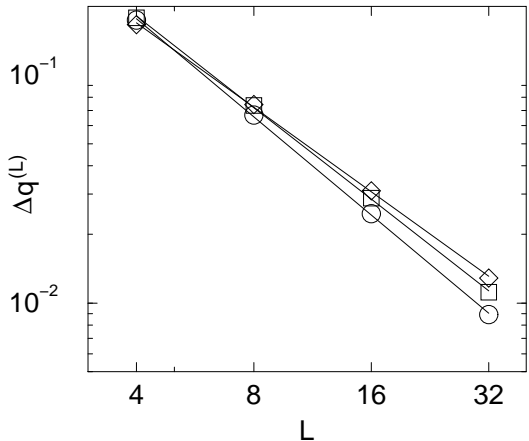}}}
\narrowtext{\caption{Width of $P(q^{(L)})$, $\Delta q^{(L)}$, as a function of
$L$ for $T=T_c$ (circles),  $T=0.7T_c$ (squares) and $T=0.45T_c$ (diamonds),
together with a power-law fit.  The exponents of the three fits are 1.43, 1.33
and 1.23.
}
\label{fig6} }
\end{figure}

Finally, let us study the width
$$\Delta q^{(L)} = \sqrt{\int_{-1}^1 (q^{(L)}-\bar q^{(L)})^2 P(q^{(L)}) 
dq^{(L)}}$$
of $P(q^{(L)})$. 
Fig.~\ref{fig6} shows our results on a double logarithmic plot, together with a 
power-law fit $q^{(L)} \sim L^{-\omega}$.
At $T=T_c$, the exponent $\omega$ is sufficiently close to $d/2=1.5$
to suggest that the leading contribution to $\Delta q^{(L)}$ comes
from the superposition of independent contributions of the different
parts of the system, just as it does for higher temperatures. Below,
we shall see explicitely that critical point nonanalyticities are
indeed subleading to the regular contributions. 

Within the framework of the droplet picture, the value of $\omega$ at
the zero temperature fixed point can be calculated as follows: The main 
excitations at low temperatures or large scales are droplets of 
flipped spins of a radius $r \simeq L$ in one of the replicas. 
Such droplets occur with a probability proportional to
$$(L/r)^d r^{-\theta} kT\,,$$
and each of them makes a contribution of the order 
$$r^{2d_s} L^{-2d}$$ to $(\Delta q^{(L)})^2$. $d_s$ is the fractal
dimension of the droplet surface and is $d-1$ in MKA, and $\theta$ is
the scaling exponent of the coupling strength and has a value around
0.24 in MKA in $d=3$. We therefore find 
$$\Delta q^{(L)} \sim \sqrt{kT} L^{d_s-d-\theta/2}\,,$$ giving $\omega
\simeq 1.12$ in MKA. Our above data for $T=0.45T_c$ are not far from
this result. For larger system size, they must ultimately converge to
it. Just as in the case of the Parisi overlap \cite{moo98}, the
crossover from critical to low-temperature behaviour is so slow that
even for $T \simeq 0.4T_c$ the asymptotic regime is not reached for
system sizes up to 32, and  the curves appear to show effective
exponents. 

For a cubic lattice, we have $d_s \simeq 2.2$ and $\theta \simeq 0.2$,
predicting a value $\omega \simeq 0.9$ within the framework of the
droplet picture. This has to be compared to the RSB scenario, where
$\omega=0$. In \cite{mar99}, it is claimed that Monte-Carlo simulation
data at $T \simeq 0.6 T_c$ and $L \le 12$ show already the signatures
of RSB. However, as yet there is no published data for $\Delta
q^{(L)}$. Only if a value of $\omega$ smaller than 0.9 is found, one
can conclude that the droplet picture is inappropriate and that RSB
occurs. As long as $\omega$ appears to be larger than 0.9, the data
are compatible with the droplet picture and are affected by finite-size effects.

\section{The link overlap in the presence of a coupling between the two 
replicas}

In \cite{mar98,mar99}, the authors suggested studying the expectation value
of the link overlap in the presence of a coupling between the two
replicas, $q^{(L)}(\epsilon)$, in order to test whether a system shows
RSB. A positive coupling $\epsilon$ in Eq.~(\ref{H}) favours a
configuration where both replicas are in the same state, while a
negative $\epsilon$ favours configurations with smaller overlaps. If
the RSB scenario were correct, the distribution $P(q^{(L)})$ would have
a finite width even for $L\to \infty$ and range from some minimum
value $q_{\rm min}$ to a maximum value $q_{\rm max}$. Consequently,
the expectation value of the link overlap, $q^{(L)}(\epsilon)$, would
have a jump from $q_{\rm min}$ to $q_{\rm max}$ at $\epsilon=0$ in the
thermodynamic limit $L\to \infty$ \cite{mar99}. In contrast, as we
will show next, the droplet picture predicts a continuous and at
$\epsilon=0$ nonanalytic function $q^{(L)}(\epsilon)$
\cite{reply}. Part of these results were published in \cite{reply}.

Within the droplet picture, the scaling dimension of $\epsilon$ in the
spin glass phase can easily be obtained. At length scale 1, $\epsilon$
is equivalent to the energy cost of flipping one spin in one of the
two replicas. On a scale $l$, this becomes the energy cost of flipping
a droplet of radius $l$ in one of the replicas, which is proportional
to $\epsilon l^{d_s}$. The scaling dimension of $\epsilon$ is
therefore $d_s$, where $d_s$ is the fractal dimension of the droplet
surface. Equivalently, $d_s$ is the fractal dimension of a domain
wall. Within MKA, we have $d_s=d-1$. The same value $d-1$ for the
scaling dimension of $\epsilon$ is also obtained from an analytical
calculation of the recursion relations for $\epsilon$ and the strength
$J$ of the random couplings near the $T=0$ fixed point. 

The positive dimension of $\epsilon$ implies that the coupling between
the two replicas is a relevant perturbation and that on large scales a
behaviour different from that of an independent system can be
seen. When $\epsilon$ is positive, the energy cost $\epsilon
l^{d_s}$ for the excitation of droplets of radius $l$ leads to the
suppression of droplets larger than $$l^* \sim
(kT/\epsilon)^{1/d_s}.$$ On scales beyond $l^*$, droplet excitations
must occur in both replicas simultaneously. This costs twice the
energy of a single droplet in an independent system. The coupled
system thus behaves on large scales exactly like a single system with
twice the coupling strength $J$. 

As stated in the preceding section, for $\epsilon=0$ droplets of size $l$ 
occur 
with a probability proportional to $(L/l)^d l^{-\theta} kT$ in a system of size 
$L$. 
Since for positive $\epsilon$ droplets
of size greater than $l^*$ are suppressed, the change 
in the link overlap due to a small positive $\epsilon$ can be
written as
\begin{equation} 
q^{(L)}(\epsilon)-q^{(L)}(0) \sim kT \sum_{l>l^*}^{L} l^{d_s-d-\theta}.
\end{equation}
In order to suppress each droplet only once,
the sum must be taken over distinct length scales $l$, e.g., $l = l^*,
2l^*, 4l^*, ...$, and it is proportional to the first
term. Therefore, we can write $q^{(L)}(\epsilon)-q^{(L)}(0) \sim kT
(l^*)^{d_s-d-\theta}$, and using the expression for $l^*$ given above
we have,
\begin{equation} 
q^{(L)}(\epsilon)-q^{(L)}(0) \sim  kT (\epsilon/kT)^{(d+\theta-d_s)/d_s}. 
\label{pose}
\end{equation}

For negative $\epsilon$, flipping a droplet of radius $l$ in one of
the replicas changes the system's energy by an amount proportional to
$l^\theta-|\epsilon| l^{d_s}$, which is negative for $l > l_c$ with
$l_c \sim |\epsilon|^{1/(\theta-d_s)}$. Therefore, there is a
proliferation of droplets beyond this length scale and the spin glass
state is completely restructured. We followed the flow of
the parameters $\epsilon$, $J$, and $\Delta \epsilon$ (the width of
the distribution of $\epsilon$) under a change of scale in the MKA and 
found
that $\Delta \epsilon$ diverges, while $J$ and $\epsilon$ eventually
decrease to zero. Such a system is an Edwards-Anderson spin glass with
the effective spins $\rho_i=\sigma_i \tau_i$.

Since droplets of size larger than $l_c$ proliferate for negative
$\epsilon$, the change in the link overlap is  given in this case by 
\begin{equation} 
q^{(L)}(\epsilon)-q^{(L)}(0) \sim -(l_c)^{d_s-d} \sim
-|\epsilon|^{(d-d_s)/(d_s-\theta)}. \label{nege}
\end{equation}

We thus find that $q^{(L)}(\epsilon)-q^{(L)}(0)$ has the form
$A_{\pm}|\epsilon|^{\lambda_\pm}$, with values $A$ and $\lambda$ that
depend on the sign of $\epsilon$. Within MKA, it is $\lambda_+ \simeq
0.62$, and $\lambda_- \simeq 0.57$. For a cubic lattice, $\lambda_+ \simeq
0.45$, and $\lambda_- \simeq 0.40$. 

\begin{figure}
\epsfysize=0.75\columnwidth{{\epsfbox{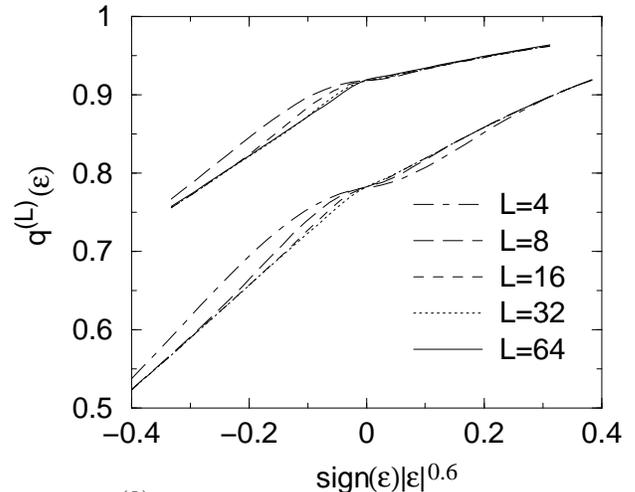}}}
\narrowtext{\caption{$q^{(L)}(\epsilon)$ in MKA at $T\simeq 0.38 T_c$ (bottom) 
and $0.14T_c$ (top) as
function of sign$(\epsilon)$$|\epsilon|^{0.6}$ for various system
sizes. Each curve is averaged over several 1000 samples.}\label{qle}}
\end{figure}

For finite temperatures and small systems, there are corrections to
this asymptotic behaviour due to finite-size effects which replace the
nonanalyticity at $\epsilon=0$ with a linear behaviour for small
$|\epsilon|$, and due to the influence of the critical fixed point,
where the leading behaviour is linear in $\epsilon$ (see below). As we have 
argued
in \cite{moo98}, the influence of the critical fixed point changes the
apparent value of the low-temperature exponents for the system sizes
studied in Monte-Carlo simulations and the MKA.  The data shown in \cite{mar99}
with an apparent value of 0.5 for ${\lambda_\pm}$ are fully compatible
with the above predictions of the droplet picture. There is no indication of a 
jump at $\epsilon=0$ in $q^{(L)}(\epsilon)$, which would be the signature of 
RSB. For the MKA, the
apparent exponent at $0.7T_c$ is close to 1 for $L\simeq 16$, leading
to the ``trivial'' behaviour found in \cite{mar99}. However, at
lower temperatures, for the same small system sizes the
above-mentioned nontrivial features predicted by the droplet picture
become clearly visible, as shown in Fig.~\ref{qle}.

\begin{figure}
\epsfysize=0.75\columnwidth{{\epsfbox{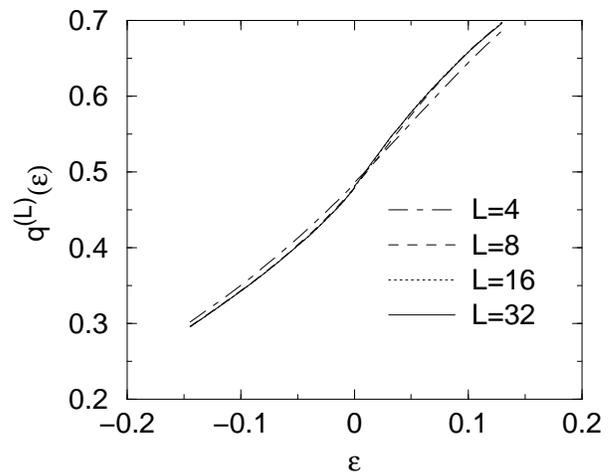}}}
\narrowtext{\caption{$q^{(L)}(\epsilon)$ in MKA at $T\simeq T_c$ for various 
system
sizes.}\label{qlcrit}}
\end{figure}

Let us conclude this section with a discussion of the link overlap at the 
critical temperature. Fig.~\ref{qlcrit} shows our result in MKA. 
Clearly, the curves are linear at $\epsilon=0$, indicating that the
regular part dominates over the singular, critical contribution. This conclusion 
is confirmed by studying the scaling dimension of $\epsilon$ at $T_c$. Iterating 
the recursion relations for the coupling constants, we find that under a change 
in length scale, $x \to x/b$, we obtain $\epsilon \to b^\phi \epsilon$ with $\phi 
\sim 1.14$. Now, $q^{(L)}(\epsilon)$ can be obtained from the free energy via 
the relation 
$$q^{(L)}(\epsilon) = (1/3N) \partial (\ln Z) / \partial \epsilon,$$
implying a scaling behaviour $q^{(L)} \to b^{d-\phi} q^{(L)}$. Substituting $b$ 
with $\epsilon$ gives then the relation 
$$q^{(L)} \sim \epsilon^{(d-\phi)/\phi} \simeq \epsilon^{1.6}.$$
Compared to the linear regular part, this singular dependence cannot
be seen for small $\epsilon$. 

\section{Finite size effects}

\begin{figure}
\centerline{
\epsfysize=0.7\columnwidth{\epsfbox{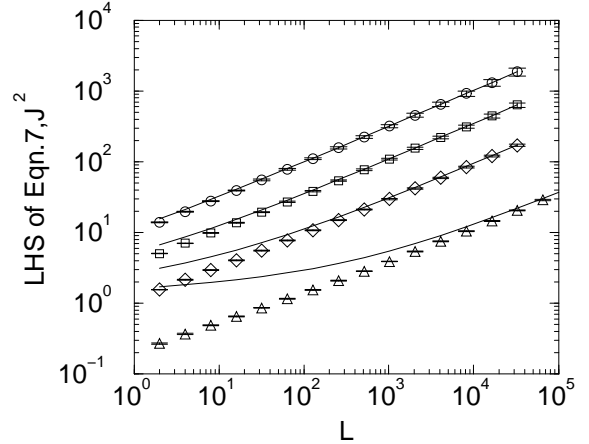}}}
\narrowtext{\caption{$J_n^2$ (solid lines) and left hand side of Eqn.7 
(symbols) as
a function of system size $L=2^n$ for (from top to bottom) $T=0.3T_c,
0.5T_c, 0.7T_c, 0.9T_c$.} 
\label{fig9} 
}
\end{figure}

As we have argued throughout this paper and in earlier work
\cite{moo98} \cite{reply}, finite size effects appear to be large for
the Ising spin glass in three dimensions. To understand this
behaviour, we iterated the MK recursion relations starting at various
temperatures below $T_c$. In Fig.\ref{fig9} the solid lines show the
data for $J_n^2$ as a function $L=2^n$ where $n$ is the number of
iterations of the MK recursion relations.  For large L, one expects
$J_n^2 \simeq 2^{2n\theta}$.  However, it is apparent that even at
$T\simeq0.7T_c$ one needs system sizes $L \simeq 100$ to see this
behaviour. Because the change in the slope $d\ln J^2 /d \ln L$ is so
slow, $J^2(L)$ appears to be described by a power law with some
effective exponent over small windows of one decade in $L$, just as we
found for $P^{(P)}(0)$ in \cite{moo98}.  To understand the large
crossover regime, we consider an expansion around the zero-temperature fixed 
point, where  the effective
temperature $T=1/J$ at a length scale $L$ can be written as 
\begin{equation}
dT/d \ln L = -\theta T + AT^3 + BT^5 + ..., 
\label{mc}
\end{equation}
where A, B, ... are constants and even order terms are 
absent because $T\to -T$ (or $\{J_{ij}\}\to \{-J_{ij}\}$) is a symmetry of 
the Hamiltonian. 
Now, $d=3$ is close to the lower
critical dimension for Ising spin glasses (so that  $T_c$ is in
some sense small), and one might expect that
truncating the above equation after the first few terms
will give a good approximation for $T$
or equivalently $J$ throughout the low temperature phase. 
We now show that this is indeed the case: keeping terms 
up to $T^5$ gives a good
description of the MK data of Fig.\ref{fig9}. 

The analysis is as follows: At $T_c$, $dT/dL = 0$ so that 
$\theta = A{T_c}^2 +B{T_c}^4$. For small deviations away from $T_c$ i.e. 
$T=T_c+\delta T$, the correlation length exponent $\nu$ is defined via
$d \ln (\delta T)/d\ln L = 1/\nu$, leading to $1/\nu = 2\theta+
2BT_c^4$. Solving for $A$ and $B$ in terms of $\nu$, $\theta$
and $T_c$, we find $A=(2\theta-1/2\nu)/T_c^2$ and
$B=(1/\nu-2\theta)/2T_c^4$. 
Now we substitute these expressions into the above equation and
integrate from a length scale $L_0$ (temperature $T_0$) to a length
scale $L$ (temperature $T_L$) to get an equation relating $T_L$
to $T_0$. Then using $T_L=1/J_L$ gives the corresponding equation for $J_L$.
Setting  $L/L_0=2^n$ and  $2\theta\nu=x$, we find
\begin{equation}
{(J_n^2-J_c^2)^x
\over{\left[J_n^2-{x-1\over x}J_c^2\right]
^{(x-1)}}} =
{(J_0^2-J_c^2)^x
\over{\left[J_0^2-{x-1\over x}J_c^2\right]
^{(x-1)}}} 2^{2n\theta},
\label{jn}
\end{equation}
where $J_0$ is the coupling at the starting point $n=0$ and
$J_n$ is the coupling after $n$ iterations. 

To test whether Eq.~(\ref{jn}) is a good approximation, we have
inserted the values $J_n^2$ obtained from the MK recursion relations
into the left hand side of Eq.~(\ref{jn}).  We have chosen $\nu=2.8$
and $\theta=0.25$, in agreement with \cite{southern77,moo98}. The
result are the data points given in Fig.~\ref{fig9}. They satisfy a
power law with exponent $2\theta$ and a prefactor $(J_0^2-J_c^2)^x
/(J_0^2-(x-1/ x)J_c^2)^{(x-1)}$, showing that Eq.~(\ref{jn}) is indeed
an appropriate description of the growth of the coupling throughout
the low temperature phase. Now, for large $J_n$ and $J_0$,
Eq.~(\ref{jn}) reduces to pure power law behaviour viz.  $J_n^2\simeq
J_0^2 2^{2n\theta}$ and the crossover length for the different
temperatures can be read off as the length for which $
(J_n^2-J_c^2)^{2\theta\nu}/(J_n^2-(2\theta\nu-1)J_c^2/2\theta\nu)^
{(2\theta\nu-1)}$ becomes of the same order as $J_n^2$. Beyond the crossover 
length, we should see the correct low-temperature exponents. 

Thus, we have seen that the three-dimensional spin glass in MKA can be
well described by an approximation that is valid close to the lower
critical dimension and that shows explicitely that crossover scales
are large.
The analysis described above is quite
general and should be applicable also to the Ising spin glass on a cubic
lattice. There is one caveat however: the analysis we have carried out
is valid for the couplings alone. Other quantities (for example
$P(q^{(P)})$ or $P(q^{(L)})$) would have to be studied separately and
it is possible that the crossover lengths would be somewhat 
different for
different quantities.

\section{Conclusion}

In this paper, we have studied the link overlap between two identical
replicas of a three-dimensional Ising spin glass in
Migdal-Kadanoff approximation.  The width of the link overlap
distribution decreases to zero with increasing system size at $T_c$ as
well as in the low-temperature phase. These findings are in agreement
with the predictions of the droplet picture. For system sizes similar
to the ones used in Monte Carlo simulations of a cubic lattice, we
find the same large sample-to-sample fluctuations and asymmetric curve
shapes as reported from those simulations. They must be interpreted as
a finite-size effect and cannot be taken as an indicator for
replica-symmetry breaking. The only reliable indicator for RSB would
be a width of the overlap distribution that does not decrease with
increasing system size. 

Similarly, the link overlap in the presence of a weak coupling between
the two replicas shows in MKA in the spin glass phase the singular
behaviour predicted by the droplet picture. Data from Monte Carlo
simulations are also in full agreement with the droplet picture. The
RSB picture predicts a jump in the mean value of the link overlap at
zero coupling strength that is not visible in Monte Carlo simulation
data published so far.

We have reproduced phenomenologically the influence of the critical point  on
the growth of the coupling constants using an approximation that is valid close 
to the lower critical dimension. This gives us a direct estimate of the
lengthscale at any given temperature beyond which one needs to go in order to
see zero-temperature (droplet) scaling without crossover effects from critical
point behaviour intruding too strongly. This crossover effect often seems to get
overlooked in the literature. For example Komori, Yoshino and Takayama
\cite{4143} in a numerical simulation of the Ising spin glass at a temperature
of $0.84T_c$ found that critical scaling of the dynamics worked well: they
found that correlation data at time $t$ could be collapsed for systems of
linear dimension $L$ for values of $L$ up to 7, by plotting against the
variable $t/L^{z(T)}$, where $z(T)$ is similar to the critical point dynamical
exponent. This behaviour was interpreted by Marinari, Parisi and Ruiz-Lorenzo
\cite{4321} as evidence against droplet scaling but it is clear from our work
that for the system sizes studied at temperatures so close to $T_c$ droplet
scaling would be quite unobservable and that critical point scaling should
indeed work quite well.
 
We therefore conclude that the droplet picture, combined with
finite-size effects, can fully explain all data for the link overlap
in the Ising spin glass. There is no evidence for the presence of RSB. 

\acknowledgements 
This work was supported by EPSRC Grants GR/K79307 and GR/L38578.

\end{multicols} 
\end{document}